# Situations d'apprentissage collectives instrumentées : étude de pratiques dans l'enseignement supérieur

**Michel Christine, Garrot Elise, George Sébastien**

*Laboratoire LIESP, INSA-Lyon*
*21, avenue Jean Capelle*
*F-69621 Villeurbanne Cedex, France*
*Christine.Michel@insa-lyon.fr, Elise.Garrot@insa-lyon.fr,*
*Sebastien.George@insa-lyon.fr*

RÉSUMÉ. Actuellement, les activités collectives sont de plus en plus souvent supportées par des plates-formes éducatives qui proposent de nombreux outils de communication, de production, de partage et de management du travail collectif. Mais il n'est pas garanti que la mise à disposition des ces outils aux acteurs (concepteurs, tuteurs et apprenants) implique leur réelle utilisation. Notre travail, par une étude de terrain, dresse un bilan des caractéristiques des situations d'apprentissage collectives instrumentées (SACI) dans l'enseignement supérieur. Notre propos est de déterminer : si les SACI existent ; quelle forme elles prennent (en termes de scénario, d'outils, de type d'activité…) ; si les recommandations issues de travaux de recherche sont mises en pratique par les concepteurs pédagogiques et si les activités prescrites par ces concepteurs se déroulent comme prévu avec les apprenants et les tuteurs ? Pour répondre à ces questions, nous nous basons plus particulièrement sur des enquêtes d'usage auprès des acteurs de SACI.

MOTS-CLÉS : Analyse et évaluation des usages, Évaluation des EIAH, Apprentissage collectif, Outil de travail collaboratif/coopératif.

**1. Introduction**

Les travaux de recherche présentés dans cet article s'inscrivent dans le projet ACTEURS TICE[1] (Activités Collectives et Tutorat dans l'Enseignement Universitaire : Réalités, Scénarios et usages des TICE). Dans le cadre de ce projet nous nous intéressons plus spécifiquement à l'étude des Situations d'Apprentissage Collectives Instrumentées (SACI) dans l'enseignement supérieur [BOURRIQUEN et al. 06]. En effet, l'apprentissage collectif est utilisé pour favoriser l'apprentissage individuel à partir d'interactions entre apprenants [DILLENBOURG 99] [DOISE et MUGNY 84]. Dans ce contexte, des activités variées sont mises en place, telles que l'apprentissage par projet, les études de cas ou les jeux d'entreprise. Elles sont de plus en plus souvent réalisées sous une forme instrumentée pour pouvoir être utilisées de manière plus flexible (en particulier à distance). Parallèlement, des pressions extérieures (académique, économique, sociale) encouragent l'utilisation des Technologies de l'Information et de la Communication (TIC). En considérant l'importance actuelle des TIC en éducation et leur usage dans les activités collectives, nous avons décidé d'observer ce que nous nommons les SACI dans le contexte de l'enseignement supérieur. Nous tenons néanmoins à préciser que la finalité de cette recherche n'est pas de comprendre les mécanismes collectifs pour améliorer la conception de système d'apprentissage comme c'est le cas dans le champs des CSCL (Computer Support for Collaborative Learning) [BANNON 89] [ISLS 06] [JERMANN 01]. Notre propos est de montrer, dans une démarche globale et interdisciplinaire, (1) les formes que peuvent prendre les SACI et leur intégration dans les programmes de formation (2) le rôle et ressenti des acteurs des SACI et (3) l'intentionnalité exprimée *a priori* par les concepteurs sous la forme de scénarios. Cette étude est réalisée par une observation et caractérisation des situations effectives sur le terrain. Après avoir défini plus précisément le terme SACI et la méthodologie d'observation, nous présentons les résultats obtenus.

**2. Problématique**

Une SACI est définie comme une situation pédagogique avec un objectif d'apprentissage (de connaissances et/ou de compétences), des acteurs identifiés, une durée et un mode d'évaluation des apprenants [BOURRIQUEN et al. 06]. Elle prend la forme d'une unité d'apprentissage scénarisée dans laquelle la production individuelle et/ou collective attendue est liée à une activité collective instrumentée par des artefacts informatiques. Dans cette définition, le terme *« collectif »* est

---

employé pour désigner à la fois les concepts *« collaboratif »* et *« coopératif »* [GEORGE 01], les sous-buts étant identiques dans une activité collaborative alors qu'il y a un partage de tâches entre acteurs dans une activité coopérative. Le terme *« production »* désigne tout résultat tangible, matérialisé et observable *a posteriori* d'une activité. Par exemple la production peut-être une synthèse individuelle ou collective, des discussions provenant d'outils de communication ou de diffusion (forums, mails, blogs, chats), des traces de connexion, etc. Il est à noter qu'une SACI peut avoir différents niveaux de granularité – elle peut ainsi être une séance, une séquence ou une unité d'enseignement – dès lors qu'elle est délimitée par les caractéristiques énoncées dans sa définition (sa durée, son objectif d'apprentissage…).

Les plates-formes éducatives proposent de nombreux outils : *outils de communication* (chat, mail, forum, vidéoconférence, audioconférence), *outils de production et de partage* (partage d'applications, tableau blanc, éditeur de texte partagé, wiki, blog, portfolio) et *outils de management du travail collaboratif* (gestion de document, outil de planification). Mais la mise à disposition de ces outils aux acteurs (tuteurs et apprenants) implique-t-elle leur utilisation ? Et leur utilisation implique-t-elle forcément une réelle activité collective entre apprenants ?

Dans cette étude, nous cherchons à voir s'il y a une différence entre ce qui est prescrit par un concepteur pédagogique de SACI et le déroulement effectif de cette même SACI. Pour identifier cette différence nous observons les intentions des concepteurs (présence d'un scénario, désignation d'outils) et les pratiques des tuteurs et apprenants. Les résultats attendus de nos travaux sont : (a) observer dans quelle proportion les SACI, telles que nous les définissons, existent réellement ; (b) observer dans les SACI identifiées la réalité des pratiques des acteurs afin d'évaluer les éventuels écarts entre les recommandations proposés par le monde de la recherche et les réalisations effectives ; (c) observer dans les SACI identifiées les écarts existant entre les activités prescrites par les concepteurs pédagogiques et le déroulement effectifs de celles-ci ; (d) apporter des hypothèses pour expliquer les différences observées au niveau des deux points précédents ; (e) ouvrir des pistes de réflexion et des ébauches de réponses pour rendre la réalisation d'activités collectives effective.

Cette étude s'inscrit autour de l'usage des outils TIC. Nous observons par exemple si les concepteurs prévoient dans la conception d'activités collectives l'usage d'outils particuliers pour effectivement favoriser la collaboration entre apprenants, si les tuteurs et apprenants les utilisent ou en utilisent d'autres de leur propre initiative ou bien encore en détournent l'usage [RABARDEL 95].

## 3. Démarche et méthodologie

L'observation a été réalisée à partir d'un guide d'entretien, d'un questionnaire préalable et d'un questionnaire final. Le *questionnaire préalable* et *l'entretien semi-directif* ont été réalisés avec un concepteur du campus numérique FORSE

(http://www.sciencedu.org/) et deux tuteurs du campus Spiral (http://spiral.univ-lyon1.fr). Les résultats ont permis d'affiner la définition d'une SACI en précisant ses caractéristiques spécifiques comme le niveau de granularité (séance, séquence, unité d'enseignement, etc.). Ils ont également servi à définir plus finement le questionnaire final de caractérisation des SACI et à localiser, parmi les 13 terrains d'observation du groupe ACTEUR TICE ceux intégrant des SACI dans leur dispositif de formation. Cette enquête préalable a permis de montrer que les SACI n'étaient pas aussi développées dans le milieu de l'enseignement supérieur que nous l'avions pensé, seules 12 SACI (alors que nous en espérions entre 30 et 40) ont été en effet identifiées. Le *questionnaire final* a été utilisé pour réaliser 18 entretiens auprès des acteurs des 12 SACI retenus. Nous nous sommes attachés à questionner ces acteurs sur le contexte global dans lequel s'inscrit la SACI (contenu, public, diplôme, discipline, genèse de la SACI) et, de manière plus spécifique, sur le type d'instrumentation et les modalités d'interactions collectives mis en œuvre dans la SACI. Plus précisément, nous les avons interrogé sur les outils informatiques de *communication*, de *production/diffusion/partage d'information ou de ressources* et de *gestion du travail collectif* proposés et utilisés.

## 4. Résultats

Les premiers résultats sur le profil global des SACI observées montrent qu'elles s'inscrivent toutes (100%) dans des formations diplômantes, dans des disciplines variées (anglais, informatique-infographie, ingénierie de la formation à distance, marketing, comptabilité, gestion d'entreprise). Les *motivations* qui ont poussé à la création de ces SACI sont d'agir sur une formation existante en présence (resp. à distance) en la complétant (35% resp. 23%) ou en la remplaçant/améliorant (47% resp. 11,8%), de créer une nouvelle formation (41,2%) ou de chercher de nouveaux public (35,3%). Les expériences sont assez jeunes puisque 40% d'entre elles en sont à leur 1$^{ere}$ session et que 70% ont été réalisées sur moins de 4 sessions. La *granularité* est très variable ; on observe en effet des SACI ponctuelles (de 2h, de 2 jours) ou des SACI sur de plus longues périodes (1 semestre à 1 an). La part d'*enseignement en présence* est faible (inférieure à 10% du temps d'apprentissage dans environ 60% des cas). Les *objectifs* sont assez variés, allant de réalisation de tâches simples (apprendre du vocabulaire, apprendre la manipulation d'un logiciel de conception graphique), à des tâches plus évoluées d'analyse globale et de gestion (de ressources, de projet, d'hommes). Il n'y a pas de *type d'activité* privilégié puisque nous relevons approximativement autant de situations d'apprentissage concernant de la recherche d'informations, du débat, de l'écriture collaborative, du projet, de l'étude de cas et de la résolution de problèmes. Les activités sont même souvent combinées : écriture collaborative dans un projet, recherche d'information et résolution de problème, débat et écriture collaborative par exemple. Le travail est prévu par le concepteur pour être *collectif* dans 64,38% des cas, ce qui est généralement suivi dans la pratique (60,71%). Le travail est plutôt de nature coopérative que collaborative (60% en effectif) mais ce chiffre est à prendre avec précaution. Il semblerait que les acteurs aient eu du mal à distinguer la collaboration de la coopération, même si une définition était donnée. En effet, nous avons observé par exemple des réponses contradictoires pour une même SACI comme : le concepteur indique une proportion d'activités collaboratives de 100% et le tuteur de 0%. Au niveau de la temporalité, les activités *synchrones* sont très souvent prescrites par le concepteur (87,5% pour les apprenants et 100% pour les tuteurs). Si les activités *asynchrones* sont un peu moins prescrites (avec 68,75% et 64,29% pour les mêmes populations) elles sont bien suivies (62,50% des SACI les voient utilisées « souvent » contre 6,25% « pas souvent ») alors que cette tendance est un peu moins forte pour les activités *synchrones* (56% des SACI les voient utilisées « souvent » contre 31,25% « pas souvent »).

L'étude spécifique de l'utilisation des outils d'interactions collaboratives, détaillée dans [MICHEL et al. 07], montre que les *outils de communication* les plus fournis sur les plates-formes sont le mail (68,75% pour les apprenants et 76,47% pour les tuteurs), le forum (68,75% pour les apprenants et 64,7% pour les tuteurs) et le chat (56,25% pour les apprenants et 58,82% pour les tuteurs). Ils sont généralement bien utilisés. Peu de plates-formes fournissent des outils de visio ou audio conférence et quand c'est le cas ils sont peu utilisés. Les apprenants utilisent plus les forums que le mail alors que les tuteurs préfèrent le mail. Le téléphone est rarement proposé aux apprenants et un peu aux tuteurs qui l'utilisent entre eux. Les chats sont assez peu utilisés mais paradoxalement, dans certaines SACI et pour les étudiants, ils le sont alors qu'ils ne sont pas proposés. Cela montre que ces acteurs ont tendance à utiliser d'autres outils de communication rapide que ceux proposés, pour des raisons de confidentialité ou d'utilisabilité (des chats externes au dispositif par exemple). La plupart du temps, les SACI observées ne proposent pas d'*outils de partage* et de *production* de ressources. Le wiki et le blog ne sont jamais utilisés. Le blog est certes rarement proposé (6,25% pour les apprenants et les tuteurs), mais le wiki l'est (50% pour les tuteurs). Ce résultat est surprenant car ces outils sont sensés être d'usage familier ; les tuteurs n'éprouvent peut-être pas le besoin de formaliser sous cette forme leur expérience. L'éditeur de texte partagé et le tableau blanc sont plutôt souvent proposés et bien utilisés alors que le partage d'application non. Cette observation nous laisse à penser que le « bon » usage ou l'usage effectif des outils de partage tient au fait qu'ils sont intimement liés à la réalisation d'une activité spécifique dans la SACI. Les outils de *management du travail collectif* sont assez peu proposés. Quant ils le sont c'est sous la forme d'outils de gestion de document, d'outils spécifiques et d'outils de planning et dans une faible mesure d'outils d'*awareness*. Les apprenants utilisent souvent les outils de planning, les tuteurs utilisent assez bien tous les outils et même certains qui ne sont pas a priori prescrits, comme les outils d'*awareness* et les outils spécifiques. Nous pouvons supposer que les tuteurs détournent l'usage initial de ces outils pour satisfaire des besoins non pris en compte par le concepteur (comme l'évaluation du travail d'un élève et de son degré d'investissement grâce aux outils d'*awareness*, la composition d'un groupe et la collaboration au sein et entre les groupes grâce aux outils spécifiques d'organisation).

## 5. Conclusion

En terme de volume, les SACI ne sont pas particulièrement développées dans l'enseignement supérieur. La motivation poussant à la création de SACI est généralement de remplacer/améliorer une formation déjà existante en présence par une formation à distance. Les expériences observées sont donc assez jeunes, n'ont pas de traits caractéristiques forts en termes de champs disciplinaire, objectif d'apprentissage, granularité ou type d'activité. Parmi l'éventail des possibilités (activités ou outils) collectives et instrumentées disponibles, très peu semblent actuellement mises en pratique de manière aboutie. Nous observons que les concepteurs qui mettent en place des situations d'apprentissage collectives ont souvent une démarche « artisanale », bien loin des recherches actuelles sur les scénarios pédagogiques et de leur standardisation. Nous ne pouvons pas déterminer si cette constatation est due à une réaction pragmatique liée au manque de moyen matériels, d'outils ou de dispositifs ; à une difficulté à mettre en application des recommandations de la recherche ou bien à une mauvaise compréhension, par le tuteur, des objectifs du concepteur ? Nos futures recherches tenteront de répondre à ces questions.

## Bibliographie